# Aristotle said "Happiness is a state of activity" - Predicting Mood through Body Sensing with Smartwatches

Gloor, P. A., Fronzetti Colladon, A., Grippa, F., Budner, P., & Eirich, J.

# Aristotle Said "Happiness is a State of Activity" - Predicting Mood Through Body Sensing With Smartwatches

P. A. Gloor, A. Fronzetti Colladon, F. Grippa, P. Budner, J. Eirich

**Abstract**

We measure and predict states of Activation and Happiness using a body sensing application connected to smartwatches. Through the sensors of commercially available smartwatches we collect individual mood states and correlate them with body sensing data such as acceleration, heart rate, light level data, and location, through the GPS sensor built into the smartphone connected to the smartwatch. We polled users on the smartwatch for seven weeks four times per day asking for their mood state. We found that both Happiness and Activation are negatively correlated with heart beats and with the levels of light. People tend to be happier when they are moving more intensely and are feeling less activated during weekends. We also found that people with a lower Conscientiousness and Neuroticism and higher Agreeableness tend to be happy more frequently. In addition, more Activation can be predicted by lower Openness to experience and higher Agreeableness and Conscientiousness. Lastly, we find that tracking people's geographical coordinates might play an important role in predicting Happiness and Activation. The methodology we propose is a first step towards building an automated mood tracking system, to be used for better teamwork and in combination with social network analysis studies.

## 1. Introduction

Frequently we are working ourselves into a rage without being aware of it; or we might get angry and become upset about our teammates without an obvious cause. What if we had an early warning system that alerts us before we are getting all stressed out, so we can calm down by taking a break, or taking a walk in the park? What if companies could offer employees effective non-monetary incentives to enroll into wellness programs? Today an increasing number of employers provide financial incentives to lose weight or help address other health problems, though only a small percentage of people decide to partner with their company and share their personal data (Cawley and Price 2013,

Cazier, Shao et al. 2007, Sarowar Sattar, Li et al. 2013).

Based on the premise that your body tells you how happy or unhappy you are, we propose a body-sensing system that automatically recognizes individual mood state and proposes corrective action. Using commercially available Pebble smartwatches we built a body sensing system that can measure individual mood states and interactions between people. In this paper we focus on exploring the connection between the states of Happiness and Activation and metrics directly collected via wearable watches.

Through the development and application of an app called Happimeter we explored the associations between two emotional states, feeling happy and feeling activated or aroused, and smartwatch-based sensors data. Psychologists often refer to happiness as positive affect, a mood or emotional state which is brought about by generally positive thoughts and feelings (Batson, Shaw et al. 1992, Beedie, Terry et al. 2005, Lu 2001a).

The goal of this study is to test the predictive power of sensor-based variables such as average heartbeat, light level, acceleration, and GPS coordinates with regards to positive mood states and happiness. In order to calibrate the sensor-based variables, we built an application that would ask smartwatch-wearer four to seven times a day to respond to a short, one-item question to record their Happiness and Activation levels (Russell 1980).

## 2. How is Happiness Defined?

Research in positive psychology (Seligman 2004) defines happiness as the frequent presence of positive emotions such a joy, interest, and pride, and the infrequent presence (although not absence) of negative emotions such as sadness, anxiety and anger. According to the OECD Guidelines on Measuring Subjective Well-being (OECD 2013), "Subjective well-being encompasses three different aspects: cognitive evaluations of one's life, positive emotions (joy, pride), and negative ones (pain, anger, worry)". Happiness is often described as a psychological state following the gratification of some important human needs or desires and it is operationalized in terms of positive affect, life satisfaction, and absence of negative affect (Diener and Scollon 2014, Lu 2001b).

Happiness is influenced on three levels, the genetic, the political, and the individual level. Part of an individual's happiness is explained by his or her genes, and it is something the individual cannot do much about it. The second level is the political context one is living in. The individual can influence it, but it takes a long time for these external variables to change. When trying to explain happiness statistically, the following environment variables have been found to be good predictors of self-reported happiness: GDP per capita, healthy life expectancy at birth, social support, freedom to make life choices, and perception of corruption (World Happiness Report 2016). The third level of influencing is the personal level, where the individual can take action to define a context that makes her or him happier. Personal variables predictive of happiness are generosity, positive affect, and absence of negative affect (World Happiness Report 2016). While the validity of personal happiness assessment has been questioned, these personal assessments are in fact surprisingly robust. In many studies, the consistency and validity of survey answers on subjective well-being have been shown. For instance, happy people smile more during social interactions, and they are more rated as happy by friends and family members (Frey and Stutzer 2002).

In a longitudinal study of normal adult development known as the Harvard Grant Study, Vaillant (2012) addresses a very fundamental question: how can we live long and happy? For 75 years, beginning in 1938, they followed 268 Harvard undergraduate men and tracked factors such as intelligence levels, alcohol consumption, relationships, and income. The Grant Study provides strong support for the growing body of research in positive psychology that has linked social ties with longevity, lower stress levels and happiness. Despite its limitations, starting with the non-inclusion of women, the study still offers a quite comprehensive overview of the factors determining happiness. As Vaillant said: "A man could have a successful career, money and good physical health, but without supportive, loving relationships, he wouldn't be happy" (Vaillant 2012).

In their review of literature on affect in organizations Barsade and Gibson (2012) define positive affectivity as the tendency of individuals to be cheerful and to experience positive moods, such as pleasure or well-being across a variety of situations, as compared to individuals who tend to be low energy and melancholy. As there exists a ripple effect of positive emotional contagion, where group members experience improved cooperation, decreased conflict, and increase perceived task performance, knowing the mood state of one's co-workers will increase positive emotional feelings in the whole group.

## 3. How is Happiness Measured?

Measuring individual's emotional state is one of the most difficult problems in affective science (Mauss and Robinson 2009). In the World Database of Happiness (Veenhoven 2013), happiness metrics are based on responses to a Likert scored survey question like: "Taking all together, how satisfied or dissatisfied are you with your life-as-a-whole these days?" In the late 1980s, researchers at the University of Oxford devised a broad measure of personal happiness, the 29-item Oxford Happiness Inventory (OHI). An alternative scale, the Oxford Happiness Questionnaire (OHQ) was then proposed by Hills and Argyle (2002) to reduce the probability of contextual and compliant answering (Argyle 2001; Hills and Argyle 1998). OHQ consists of single items that can be answered on a six-point Likert scale and embedded into larger questionnaires. Participants are asked how much they agree or disagree on a number of statements about happiness.

In a few studies that aimed to estimate happiness using a single-item scale, researchers found that short scales were as valid as long scales, and lengthening a scale beyond some point was found to actually weaken its validity (Bell and Lumsden 1980). If researchers are primarily interested in measuring a life satisfaction score, there might be no benefit in asking respondents to address multiple questions. Measuring happiness via a single item has been demonstrated to be reliable, valid and viable (Abdel-Khalek 2006).

Based on these results, in this study we utilize a more sophisticated method, using experience-based sampling (Hulburt and Schwitzgebel 2013). At a random time per day the user is polled four to seven times on a smartwatch to rate her/his happiness. The advantage of this methodology is that we ask participants to stop at certain times and quickly report their emotional state in real time, which helps reduce biases due to wrong recollection of past events and rating of emotions, such as the fading affect bias where negative memories fades faster than affect associated with positive memories (Skowronski, Walker et al. 2014).

A recent study conducted at the MIT Media Lab tested a computer vision based system that automatically encourages, recognizes and counts smiles among passerby (Hernandez, Hoque et al. 2012). In their follow-up survey with participants, the authors found that the system made people smile more than they expected, and it made them and other passerby around them feel temporarily in a better mood. Quantitative data collected through their system indicated that people were smiling more during the weekends, during campus events and around graduation day, and less during exams.

The measurement of self-efficacy and self-perception of self-efficacy, which might lead to higher performance, suggest that similar effects might also be at play for the self-perception of happiness. For instance, the performance of girls to solve mathematical problems could be improved by nurturing self-efficacy of the girls. Bandura (1997) showed that displaying edited videotapes to study participants performing a task, where the unsuccessful attempts were edited out of the tape, was leading to actually improved performance. Seeing oneself perform successfully increases self-efficacy, and this increase in self-efficacy leads to improved real performance. Similarly, seeing where, when, and with whom one is happy, might lead the individuals to actively seek these situations, leading to actually improved happiness of the individual.

Previous studies have used traditional surveys and questionnaires to measure individual feelings, with obvious disadvantages due to costs and time involved for both the respondent and the researcher. Web-based surveys have the advantage of reducing cost of data collection as well as improving data quality. However, a big disadvantage is the possible bias due to low and selective participation. In our study we limited this risk by simplifying the questions and embedding a few of them into wearable devices, which transformed a tedious activity such as filling out a survey into a game with immediate rewards (e.g. feedback to users on activation and happiness levels). Also, surveys rely on the key assumption that people do not lie while responding to questions, which is not necessarily true especially when respondents are in front of researchers or feel they are observed (Blattman, Jamison et al. 2016). With our method, participants do not feel any peer pressure to respond in a positive way as they simply select a mood state while nobody is watching. They have full control of what is shared and have an immediate reward thanks to the reporting immediately available on the dashboard.

## 4. Happimeter: a Smartwatch-based Body Sensing System

While self-efficacy has been hyped too much (Biglan 1987), and been made its own mean to an end, it has been found, in more recent research, that causality and correlation have been mixed up. Just believing that one is good in math does not make one a math star. We speculate that there might be a similarly complex relationship, where just believing that one is happy does not make one happy. Our system is based on years of experience working with sociometric badges developed at MIT's Media Lab, which record location, speech, and energy levels of people wearing them, and also note when members of workgroups are interacting in person (Dong, Olguin-Olguin et al. 2012). These badges are a powerful tool for gathering data on workplace interactions, but are relatively expensive and can be difficult to use in long-term studies. Other researchers demonstrated the validity of using self-report survey data as a good approximation of observational data collected via mobile phones (Eagle, Pentland et al. 2009). Prior research analyzing e-mail archives (Gloor 2016) and interpersonal interactions using sociometric badges (Olguin-Olguin, Waber et al. 2009) has shown that communication patterns of individuals and teams can be calculated automatically from communication archives, and that positive mood states and particular modes of interaction are associated with higher quality teamwork (Grawitch, Munz et al. 2003). Computer recognition of mood states was also implemented considering other body signals, such as the recognition of facial expressions or human speech (Freitas, Peres et al. 2017, Hernandez-Matamoros, Bonarini, et al. 2016, Mencattini, Martinelli et al. 2014, Zhang, Mistry et al. 2016).

In this study, we developed a lightweight, inexpensive, and non-intrusive sensor system that is easy to use over extended time periods, using smartwatches instead of sociometric badges. We are integrating the smartwatch with each individual's smartphone to access the phone's location sensing and data transmission capacity, as well as its processing power. The smartwatches also provide data on lighting level and heart rate. Unlike the badges, the watches are designed to be worn constantly, naturally and non-intrusively, and their rechargeable batteries have robust charge length. Their displays also enable easy two-way communication to give status updates to wearers (Chuah, Rauschnabel et al. 2016). The smartwatch uses its built-in accelerometer, light sensor, microphone, and heart rate sensor to gather data, while location is detected by the smartphone. Data from both devices is uploaded directly to a server.

We built an app called Happimeter to collect data from the accelerometer, light sensor, microphone, and heart rate sensor in the smartwatch and location from the smartphone. The Happimeter app polls users 4-7 times per day by vibrating the smartwatch, and asks them to enter their mood states. On the smartphone the app runs in the background and continuously transfers sensor data to a server. A responsive web site shows users their sensor data. Figure 1 illustrates the dashboard providing immediate user feedback from the mood poll and the sensor readings. Based on the mood values provided by the users though the app, it reports the fluctuation in Activation and Happiness (or Pleasance), as well as heart rates in beats per minute (BPM) and an estimate of the user's personality traits when he or she took a personality test based on the International Personality Item Pool (Goldberg, Johnson et al. 2006, Johnson 2014).

Figure 2 illustrates an example of the ability of the app Happimeter to associate Happiness (increasingly brighter green) with smartwatch-wearer's geo-location.

Other researchers have been using smartphones to track mood of their owners over extended periods of time and correlated it with their location (Doherty, Lemieux et al. 2014, Sandstrom, Lathia et al. 2016). There are some potential limitations, though, in using exclusively the smartphone-embedded sensors to measure mood changes, and that is the main reason why we also included data from the smartwatch. For instance, collected locations might be inaccurate, since people might be charging their phone in a place that is different from where they currently are located (e.g. they might be in class while their smartphones are in their dorm). In addition, collecting accurate heartbeats measurements using a smartphone might introduce a bias as it might require a specific action, such as pushing a finger on the camera, which triggers a possible change in the emotional response.

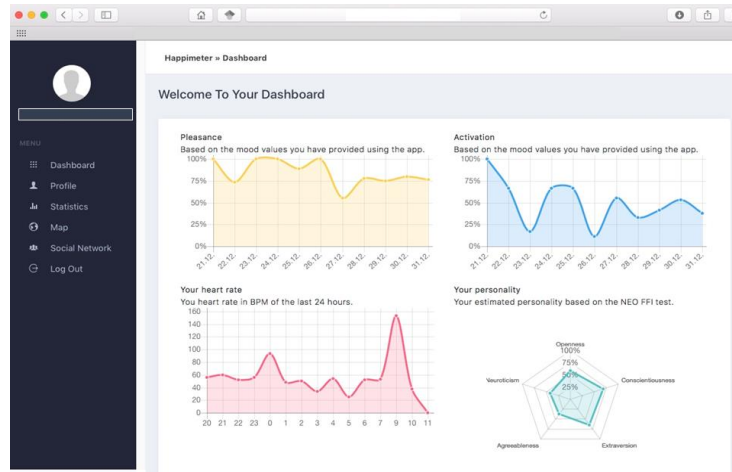

**Figure 1** Responsive web site with a dashboard presenting user feedback

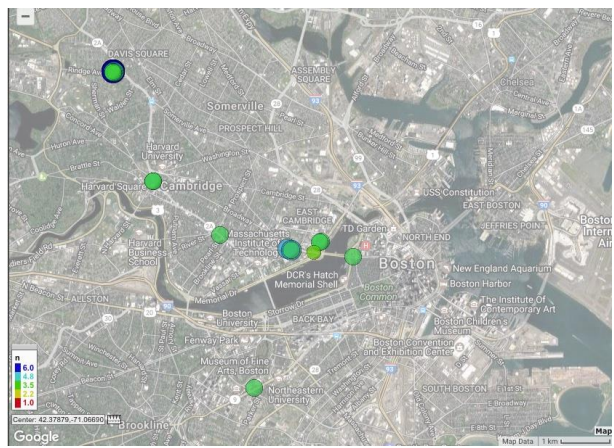

**Figure 2** Example of GPS location and mood identification via the Happimeter

## 5. Experimental Setup and Study Variables

### 5.1 Participants

The experiment involved 17 people wearing the smartwatch from December 19th, 2016 to February 3rd, 2017. Our sample included graduate students, researchers, faculty members, consultants, and business industry leaders, with age ranging from 23 to 56. Their nationalities were German, Swiss and American. When downloading the Happimeter app participants agreed to participate in this study and were sent instructions via the smartphone. Participants were polled 4-7 times per day via a vibration of the smartwatch, and were prompted to enter their mood states.

We are aware of the biases of a voluntary response sample compared to a random sample, as some members of the intended population are less likely to be included than others. At the same time, our goal was not to make inference on how the body sensors would affect those people. Our study aims at exploring a new methodology to recognize mood changes based on data recorded through smartwatches.

### 5.2 Measures

We implemented a four-outcome grid in the two dimensions "Pleasance" and "Activation" relying on the Circumplex Model of Affect (Posner, Russell, et al. 2005). In this model, the valence dimension (pleasant vs unpleasant) is on the horizontal axis and the Activation dimension (activated vs non-activated) is on the vertical axis (Posner, Russell et al. 2005, Russell 1980). Based on the two dimensions of "high Pleasance-low Pleasance" and "high Activation -low Activation", we built a system that would ask one single question "How do you feel?"; the responses would be displayed on a screen in the form of single states. Examples of emotional states connected to happiness were: feeling content, serene, calm, relaxed; the feeling of Activation was exemplified with states like feeling alert, excited, aroused (Posner, Russell et al. 2005).

Similarly to the circumplex model of affect, we assume that all emotional states can be understood as a linear combination of two dimensions, one related to valence and the other to arousal or alertness (Barrett 2006, Rafaeli, Rogers et al. 2007). Our model, described in Figure 3, reflects the assumption that specific emotions are connected to patterns of Activation within these two continua. As proposed by Posner, Russell et al. (2005) joy could be conceptualized "as an emotional state that is the product of strong Activation in the neural systems associated with positive valence or pleasure together with moderate Activation in the neural systems associated with arousal. Affective states other than joy likewise arise from the same two neurophysiological systems but differ in the degree or extent of Activation".

We collected four different mood states as combination of two levels of Pleasance and two levels of Activation. This is similar to the approach followed by LiKamWa, Liu et al. (2013) in building MoodScope, a sensor that measures the mental state of the user based on how the smartphone is used.

On our Four-outcome grid illustrated in Figure 3, we positioned Happiness on an angle very close to high Pleasance, based on the results of Russell (1980) and Posner et al (2005). As demonstrated by Russell (1980), affective space is bipolar and antonyms are positioned approximately 180°: "Beginning with happy at 7.8°, we can see that increases in angle at this point in the circle correspond to the increases in arousal and slight decreases in pleasure".

We operationalize Happiness as a binary variable equal to zero if someone is unhappy and 1 if someone is happy. Figure 3 illustrates the four-outcome grid which is a combination of the model proposed by Russell (1980) and the model proposed by Posner et al (2005), who identify happiness as aligned with high Pleasance.

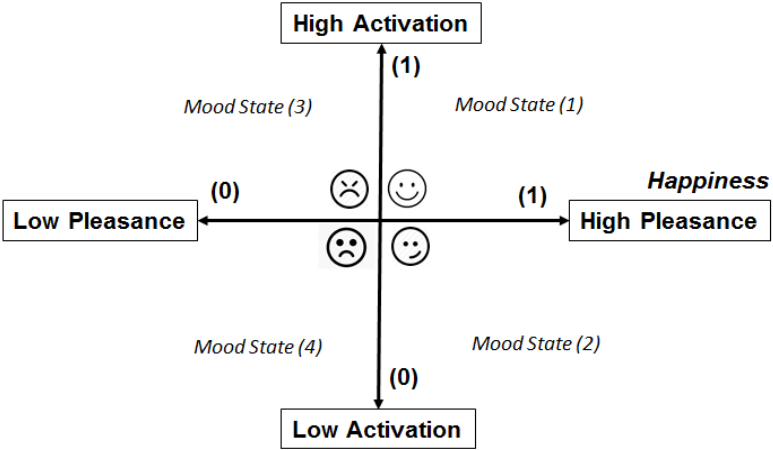

**Figure 3** Four-outcome grid used to elicit responses on mood states

Activation was operationalized as a binary variable to indicate whether individuals feel activated or inactivated. In order to test the accuracy of the machine learning algorithms that we use to assess simultaneously the dimensions of Happiness and Activation, we introduce another variable called mood state. Mood state is a categorical variable created to classify the four possible combinations of Happiness and Activation. Mood state has the value of 1 if both Happiness and Activation are 1; it has the value of 4 if both the dependent variables score 0; 3 if Happiness is 0 and Activation is 1; and 2 if Activation is 0 and Happiness is 1.

The other variables are directly recorded by the smartwatch. The light level measures the environmental light level, at the moment of measurement and it ranges from 0 to 5. BPM measures the average number of hearth beats per minute. The acceleration represents the magnitude of movement of the person in the physical space and VMC (vector magnitude counts) is a measure of the total amount of movement recorded by the smartwatch: more vigorous movement yields higher VMC values.

Our first control variable was based on the time of recording and represented the distinction between weekend and holidays. During our experiment we had two major Holidays: Christmas and New Year's Eve. Measuring the average number of heart beats per minute revealed its utility also for data cleaning purposes. In fact, it may happen that the smartwatch sensors return values that are not completely reliable – for several possible reasons, including internal malfunctioning, low battery or because the individual is not wearing the smartwatch at the moment of data collection. Considering the number of hearth bits, we filtered out all the observations where this number was zero or suspiciously low. This gave us about 17,000 useful observations.

Other control variables were age, gender, and weight (expressed in Kg). Average age was 29 years and the sample's average weight was 72 Kg (157 pounds). About 30% of the respondents were male. Gender is an interesting variable as it could provide insights into different uses of smartwatch technology, as demonstrated by a recent study on gender and the gratification of information acquisition (Zhang and Rau 2015). Sharing personal data such as weight and age could make people feel uncomfortable; therefore, we stressed repeatedly during the experiment that data was completely anonymous and was never associated to individual names in the analysis. We did not find any particular resistance from our sample in sharing their correct age or weight. This could be explained with the increasing number of applications asking for personal data in the context of independent and employer-based wellness programs.

Smartwatches were also able to collect the GPS coordinates (latitude, longitude and altitude) of the individuals wearing them at the time they answered the mood-related questions. Since sport and exercise appears to result in increased happiness (Argyle 2001; Hills and Argyle 1998) we included sportiness level among the control variable (from 1 to 3, where 1=low).

Controlling for age may offer interesting insights, since children and younger adults tend to describe their emotions solely in terms of valence (e.g. "I feel bad" or "I feel good"), lacking the nuances evident in adult descriptions (e.g. "I feel excited" or "I feel content") due to a better conceptualization of their own affective (Saarni 1999).

Past research showed that happiness can be influenced by several other factors, such as individuals' personality (Cheng and Furnham 2001, Demir and Weitekamp 2007, Tkach and Lyubomirsky 2006). Accordingly, we asked respondents to complete the Big 5 Personality Test to assess their personality profiles (McCrae and Costa 2003). We used the five factor model of personality and administered a test based on the International Personality Item Pool (Goldberg, Johnson et al. 2006). In particular, we used the 120-item version of the IPIP-NEO which covers the traditional five broad domains, namely Neuroticism, Extraversion, Conscientiousness, Agreeableness, and Openness to Experience (Johnson 2014). Recent research has demonstrated that this test is a reliable and valid measure of the five personality factors (Maples, Guan et al. 2014). Figure 4 synthesizes the variables of this study.

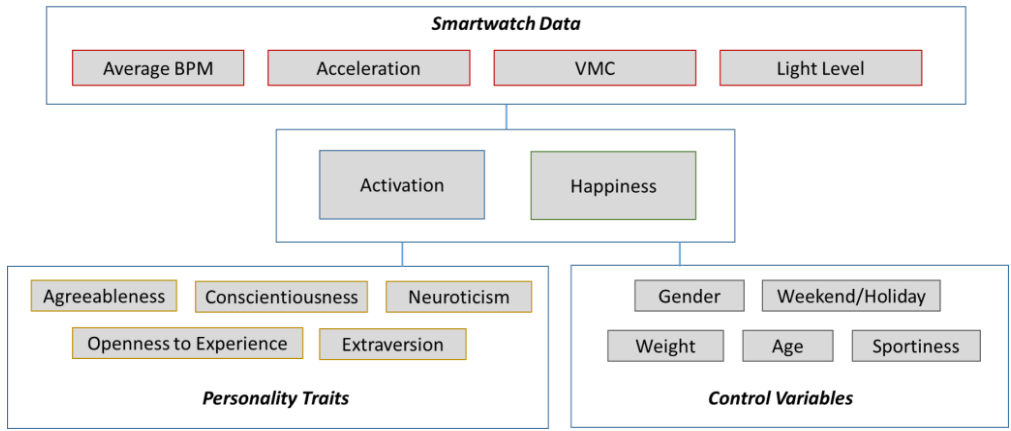

**Figure 4** Variables used in the study

## 6. Results

Both dependent variables – Happiness and Activation – have a significant negative correlation with heart beats, which may indicate that happy people are more relaxed. This seems to be aligned with studies on happiness and creativity showing that happy people are more relaxed and more open to new experiences, which boosts creativity (Pannells and Claxton 2008). Being happy and activated was negatively correlated with levels of light, due to a probable calming effect of diffused lights, since dim lighting is usually associated with relaxation, and winding down after work (Meier, Robinson et al. 2007). Looking at personality traits, a strong positive correlation was found with agreeableness, which could indicate that the more people are kind and cooperative, the more happy and activated they are. We also found a positive correlation between VMC (vector magnitude counts) and heart beats, activity and light level. This is not surprising since an increase of light in the environment could lead to more variation in body movement, higher heart beats and more activity (Meier, Robinson et al. 2007, Xu and Labroo 2014). It also seems that people tend to be happier when they are moving more intensely, though the correlation with acceleration was rather low.

Both dependent variables are highly correlated between themselves. Finally, it seems that people are feeling less activated during the weekend, which could be attributable to a more relaxed pace of life.

**Table 1** Pearson's correlation coefficients (N= 16770)

|   |             | 1       | 2        | 3       | 4     | 5 | 6 | 7 | 8 |
|---|-------------|---------|----------|---------|-------|---|---|---|---|
| 1 | Happiness   | 1.000   |          |         |       |   |   |   |   |
| 2 | Activation  | .490**  | 1.000    |         |       |   |   |   |   |
| 3 | Average BPM | -.174** | -.318**  | 1.000   |       |   |   |   |   |
| 4 | Light Level | -.111** | -.107**  | .138**  | 1.000 |   |   |   |   |

| | | | | | | | | |
|---|---|---|---|---|---|---|---|---|
| 5 | **Acceleration** | .044** | .027** | .125** | .075** | 1.000 | | |
| 6 | **VMC** | .025** | -0.014 | .431** | .179** | .272** | 1.000 | |
| 7 | **Neuroticism** | -.052** | .049** | -.066** | .022** | -0.009 | -.017* | 1.000 |
| 8 | **Extraversion** | -.087** | -.097** | -0.005 | -.028** | -.030** | -.024** | -.466** | 1.000 |
| 9 | **Openness to Experience** | -.044** | -.029** | -.029** | -.020** | -.024** | -0.008 | -.821** | .647** |
| 10 | **Agreeableness** | .258** | .260** | -.031** | -.033** | .020* | 0.005 | 0.001 | -.455** |
| 11 | **Conscientiousness** | -.089** | .044** | 0.008 | -.043** | -.051** | -.032** | -.476** | .514** |
| 12 | **Weekend/Holiday** | -0.011 | -.073** | .069** | -0.004 | .043** | .091** | -.096** | .068** |
| 13 | **Gender Male** | .062** | .169** | -.072** | -.082** | -0.001 | -.056** | .021** | 0.008 |
| 14 | **Age** | 0.010 | .103** | -.076** | -.040** | .086** | .103** | .047** | 0.004 |
| 15 | **Weight** | -0.003 | .061** | .033** | 0.008 | -.016* | -.030** | -.028** | .015* |
| 16 | **Sportiness** | .045** | .198** | -.060** | 0.004 | .036** | -0.004 | .023** | .029** |

| | | 9 | 10 | 11 | 12 | 13 | 14 | 15 | 16 |
|---|---|---|---|---|---|---|---|---|---|
| 9 | **Openness to Experience** | 1.000 | | | | | | | |
| 10 | **Agreeableness** | -.202** | 1.000 | | | | | | |
| 11 | **Conscientiousness** | .685** | -.232** | 1.000 | | | | | |
| 12 | **Weekend/Holiday** | .039** | -.105** | .023** | 1.000 | | | | |
| 13 | **Gender Male** | 0.011 | -0.007 | 0.013 | .029** | 1.000 | | | |
| 14 | **Age** | -0.006 | -.022** | -0.010 | -0.008 | .036** | 1.000 | | |
| 15 | **Weight** | 0.012 | .018* | .030** | .047** | .531** | -.153** | 1.000 | |
| 16 | **Sportiness** | .035** | -0.006 | .024** | 0.006 | .397** | .206** | .207** | 1.000 |

*p<.05; **p<.01.

Since our data represents unequally spaced repeated measures over time on a sample of 17 individuals, we extended the results deriving from the correlations and built multilevel logit models to estimate Happiness and Activation. The use of multilevel modeling for repeated measures over time is a common practice supported by past research (e.g., Hoffman and Rovine 2007, Singer and Willett 2003). In general, the fact that some measures – such as personality traits, age and gender – are time-invariant is taken into account by multilevel models: the effect of these variables is expressed by variance reductions at level 2, attributable to the differences among study participants. Indeed, the multilevel models that we present have repeated measures (level 1) nested within individuals (level 2). The residual variance is reduced by considering time-variant predictors, i.e. body signals and light level.

**Table 2a** Predicting happiness: multilevel logit models

| | Model 1 | Model 2 | Model 3 | Model 4 | Model 5 | Model 6 |
|---|---|---|---|---|---|---|

| | Model 1 | Model 2 | Model 3 | Model 4 | Model 5 | Model 6 |
|---|---|---|---|---|---|---|
| **Average BPM** | | | | | -.017** | -.018** |
| **Light Level** | | | | | -.476** | -.477** |
| **Acceleration** | | | | | 2.68e-4** | 2.32e-4** |
| **VMC** | | | | | 4.75e-6** | 4.56e-6** |
| **Neuroticism** | | | -.010** | | | -.008** |
| **Extraversion** | | | -.019** | | | |
| **Openness to Experience** | | | | .002* | | |
| **Agreeableness** | | | | .018** | | .017** |
| **Conscientiousness** | | | | -.006** | | -.010** |
| **Weekend/Holiday** | | -.072 | | | | |
| **Gender Male** | | .499 | | | | |
| **Age** | | .014 | | | | |
| **Weight** | | .0173 | | | | |
| **Sportiness** | | -.704 | | | | |
| **Constant** | 1.041** | .638 | 2.432** | .667* | 2.666** | 3.197** |
| **ICC** | 0.205 | | | | | |
| **Groups** | 17 | 17 | 17 | 17 | 17 | 17 |
| **N** | 16770 | 16770 | 16770 | 16770 | 16770 | 16770 |
| **AIC** | 19858.95 | 19860.73 | 19547.05 | 18693.56 | 19240.77 | 18021.25 |
| **BIC** | 19874.41 | 19914.82 | 19577.96 | 18732.20 | 19287.13 | 18090.79 |

*p<.05; **p<.01.

**Table 2b** Predicting activation: multilevel logit models

| | Model 1 | Model 2 | Model 3 | Model 4 | Model 5 | Model 6 |
|---|---|---|---|---|---|---|
| **Average BPM** | | | | | -.023** | -.025** |
| **Light Level** | | | | | -.524** | -.468** |
| **Acceleration** | | | | | 2.16e-4** | 2.32e-4** |
| **VMC** | | | | | 8.05e-6** | 8.67e-6** |
| **Neuroticism** | | | -.001** | | | |
| **Extraversion** | | | -.019** | | | |
| **Openness to Experience** | | | | -.007** | | -.010** |
| **Agreeableness** | | | | .022** | | .022** |
| **Conscientiousness** | | | | .027** | | .030** |
| **Weekend/Holiday** | | -.434** | | | | -.320** |
| **Gender Male** | | 1.795* | | | | 1.759** |
| **Age** | | .016 | | | | |
| **Weight** | | -.044 | | | | |
| **Sportiness** | | .920 | | | | |
| **Constant** | -.950* | -1.093 | .276 | -3.452** | 1.023** | -2.212** |
| **ICC** | 0.422 | | | | | |
| **Groups** | 17 | 17 | 17 | 17 | 17 | 17 |
| **N** | 16770 | 16770 | 16770 | 16770 | 16770 | 16770 |
| **AIC** | 18622.92 | 18505.5 | 18381.41 | 16946.73 | 17679.08 | 15960.66 |
| **BIC** | 18638.38 | 18559.59 | 18412.31 | 16985.36 | 17725.45 | 16045.66 |

*p<.05; **p<.01.

Lastly, our choice is also aligned with previous research studying longitudinal body signals through sociometric badges – which used predictive models based on correlation, multiple linear regression and machine learning algorithms, such as support vector machines (e.g., de Montjoye, Quoidbach et al. 2013, Gloor, Oster et al. 2010). In Table 2a and Table 2b, we first present the empty model for each dependent variable, Happiness and Activation respectively; in models 2, 3 and 4 we test the effects of the control variables; in the fifth model we test the predictive power of sensor data; finally in the sixth model we combine significant predictors to obtain the best model. We only present models with random intercepts because including random slopes did not improve the fit. We chose not to include the GPS coordinates in these models, since we see no specific reason for a linear association of such coordinates with the dependent variables. We did not test all the personality traits together since we found a multicollinearity problem when including Neuroticism and Openness to Experience in the same model (they present a high negative correlation in our sample). In the case of Activation, the intraclass correlation coefficient shows that about 42% of the variance is attributable to individual differences of respondents; this value drops to about 21% in the case of Happiness.

Comparing Model 1 with Model 6, we notice a significant reduction in AIC and BIC scores, in both cases, proving the value of our predictors. Both Happiness and Activation seem to be higher when the heart beats and the light level are lower, and when acceleration and VMC are higher. However, the variables inferred from the accelerometer data (VMC and acceleration) have a very small, almost negligible, effect size. With regard to personality traits, we find that people with a lower Conscientiousness and Neuroticism and a higher agreeableness tend to be happy more frequently, which reinforces our correlation results. On the other hand, more Activation can be predicted by lower Openness to experience and higher agreeableness and Conscientiousness. Activation also seems to be lower on weekends and on average higher for the male respondents. We see no influence of age, weight and sportiness on the dependent variables.

The multilevel logit models presented in Tables 2a and 2b are important to confirm the predictive power of our sensor variables, when combined with the control variables. However, their predicted probabilities depend on random intercepts, which are different for each individual; for this reason, we decided to complete our experiment testing several machine learning algorithms, in order to obtain more accurate and generalizable classifications. Using the software Weka (Holmes, Donkin et al. 1994), we found that the classification made by means of the random forest algorithm (Liaw and Wiener 2002) produced the best results. Random forest has the advantage of reducing the problem of overfitting that can arise when using decision trees (Breiman 2001). Replicating our experiment 100 times for each dependent variable, and choosing each time a random test set made of 30% of the observations, we achieved an accuracy in classifications of 96.03%% (Cohen's Kappa = 0.92) for Activation, of 94.57% (Cohen's Kappa = 0.88) for Happiness and of 92.91% (Cohen's Kappa = 0.89) for the Mood State which represents the four possible combinations of Happiness and Activation.

|  | Happiness | | Activation | | Mood State | |
| --- | --- | --- | --- | --- | --- | --- |
|  | Accuracy | Cohen's Kappa | Accuracy | Cohen's Kappa | Accuracy | Cohen's Kappa |
| **Including GPS Data** | 94.57% | 0.88 | 96.03% | 0.92 | 92.91% | 0.89 |
| **Excluding GPS Data** | 81.10% | 0.58 | 84.26% | 0.67 | 73.91% | 0.61 |

Table 3 Accuracy of random forest classifications

These results, which are quite promising also considering the good values of the Cohen's Kappa, are dependent on the location of respondents. Removing the location parameters (latitude, longitude and altitude), the accuracy of the classification models remains good, but drops to lower values as shown in Table 3. Accordingly, we maintain the importance of tracking the exact location of respondents and of categorizing such locations in future research.

## 7. Discussion

Our results support the validity of the measurements collected through smartwatch, used to explore some boosters of happiness and activation. Our main goal was not to identify which of the several factors that can impact happiness have the most predictive value. There is an extensive literature in the area of positive psychology that aims at uncovering the boosters of happiness (Argyle 2001, Lyubomirsky, King et al. 2005). In this study, we wanted to demonstrate the opportunities available to researchers when using body sensing devices. We identified some key triggers of positive well-being and happiness that could be further explored with larger samples; these include frequent movement, heart beats, change in the room lighting, tendency to help and be kind with others, as well as specific latitude and longitude. An additional contribution of this study is to explore relationships among traditional and non-traditional happiness-related variables. For instance, the strong association between higher Agreeableness and Happiness confirms previous studies in positive psychology showing that performing acts of altruism or kindness boosts happiness. In Lyubomirsky, King et al. (2005) study, doing five kind acts a week, especially all in a single day, gave a measurable boost to happiness. Similarly, Seligman (2004) found that individuals can develop durable levels of happiness by nurturing "inherent traits" such as optimism, kindness, generosity, originality, and humor. We also found that people with low conscientiousness and low neuroticism were happier than others. This is consistent with a recent study on personality and subjective well-being which found that neuroticism was most strongly associated with scores on the Depression–Happiness Scale, a measure in which greater happiness is defined by higher scores on positive thoughts and feelings, as well as lower scores on negative thoughts and feelings (Hayes and Joseph 2003). The same study found evidence that conscientiousness was a better predictor of life-satisfaction than extraversion. Literature on personality and self-reported health also found that neuroticism and conscientiousness are associated with chronic illnesses (Goodwin and Friedman 2006) and physical health (Löckenhoff, Sutin et al. 2008).

In addition, we found that high Activation can be predicted by higher agreeableness and conscientiousness. This could be explained considering that individuals high in conscientiousness are predisposed to be organized, exacting, disciplined, dependable, methodical, which could lead to feeling more alert, watchful and attentive to environmental factors than others. On the other end, agreeable people have been found to be better able to control anger and negative affect in situations involving frustration, and they are often functional during conflict resolution; this could explain why they reported feeling excitable and alert more frequently than others (Costa, McCrae et al. 1991, Graziano, Jensen-Campbell et al. 1996). We also found that high Activation was associated with lower Openness to experience. This could be explained by the typical behaviors of open people, their tolerance of ambiguity, and preference for complexity. Open people are characterized by their rich and emotional lives, and their behavioral flexibility, which could make them comfortable to encounter uncertainty and therefore not feeling particularly activated (McCrae and Costa 1997).

While our results seem aligned with previous studies exploring the determinants of happiness, the additional value of our experiment is the application of innovative methods and tools that have important practical contributions. Today, wellness programs have become common and more than 90% of medium-large companies have adopted initiatives that offer some type of incentive, often monetary, to improve employee health and decrease employer costs associated with health insurance claims (Cawley and Price 2013). Our study offers some insights that could be used by Human Resource managers to increase the intangible benefits of these programs, using "happiness" as an additional incentive for employees. Besides earning money for reporting and sharing personal data, employees could be offered the opportunity to reflect on their own happiness level and understand which factors are mostly impacting this. Our method and application suggest a new way for employees to interpret the data they share with their company. The feeling of being more in control of the shared data and the actionable insight they receive could be an additional factor to increase the percentage of people who enroll and use the wellness programs. The control variables - i.e. age, weight and sportiness – had no influence on feeling happy or activated. The result on age is not surprising because our sample comprised individuals in their adult age, with no children or elderly people who might report their emotional states in a less sophisticated way. More surprising was the result on sportiness, since sport and exercise had been associated with increased happiness (Argyle, 2001, Hills & Argyle, 1998).

Our results on the effect of room lighting are aligned with previous research showing that light intensifies emotional response. In a recent study, Xu and Labroo (2014) demonstrated that light underlies perception of heat which in turn can trigger the hot emotional system. Therefore, turning down the light can reduce emotionality and lead to more rational decisions. Differently from what Xu and Labroo (2014) found regarding an unlikely association between bright environment and feeling awake or active, our correlations results seem to indicate that individuals who feel more relaxed and not activated are the ones who are surrounded by dim light. Figure 5 summarizes our results, with dotted lines representing weaker impact on the dependent variables.

Body area sensor networks share some of the opportunities with general wireless sensor networks. A key difference, though, is that, in order to achieve social acceptance, body area sensors have to rely on nodes that are noninvasive (Hanson, Powell et al. 2009). Scholars have been increasingly interested in the social implications of ubiquitous computing and its impact on privacy (de Montjoye, Hidalgo et al. 2013; Langheinrich 2001). In a world of smart communication and computation devices, everything we say, do, and sometimes feel, could be digitized and stored. This calls for a need for systems to be designed to require users to explicitly agree before taking privacy decreasing actions (Neustaedter and Greenberg 2003). Such studies require collection of data from the smartwatch sensors, data on communication patterns inside work groups, and performance metrics for those groups. To protect the privacy of participants, we preserved anonymity, informed participants of the data collected in the platform, and educated them on how the system worked.

In future work communication data will be collected by mining e-mail/Skype/phone archives, which our group has been doing for 15 years. Face-to-face interactions will be discerned by smartphone proximity. We can then use this sensing system to track work group mood, interactions, and quality of group work output. The long-term vision is that smartwatches can become a key element of team coordination providing valuable feedback to adjust collaboration behavior based on people's moods and the kind of tasks a work group is tackling.

Collecting and analyzing such data raises important privacy issues that need to be considered (Butler 2007; Carpenter, McLeod et al. 2016). At the same time, it is important to reflect on the benefits related to improving individuals' health, informing public health decisions and building communities.

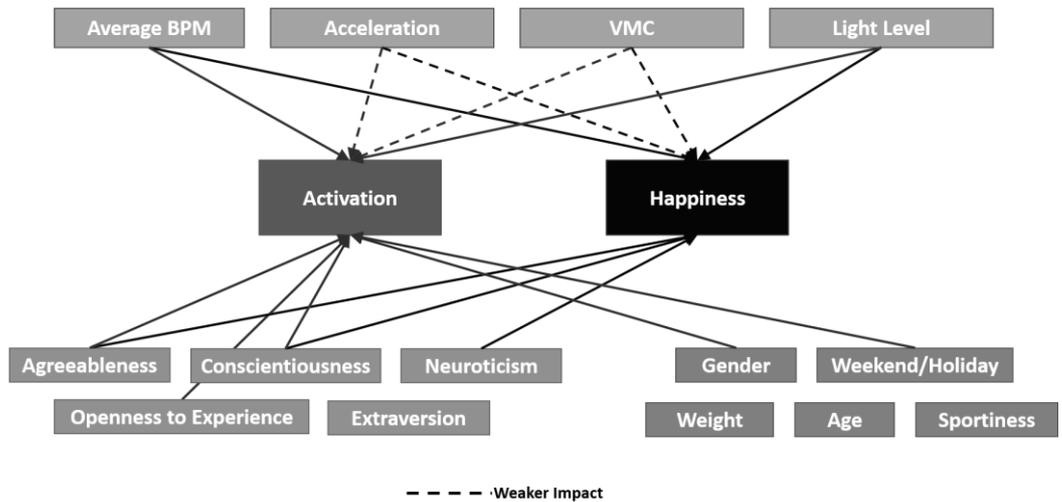

**Figure 5** Significant predictors of happiness and activation

## 8. Limitations and Future Research

In building our measurement system, we recognize the limitations of all introspective methods, which seem vulnerable to false interpretation as unreliability of memory (Hulburt and Schwitzgebel 2013). To minimize possible biases, we built a system following the approach proposed by Hulburt and Heavey (2001): in Hulburt's Descriptive Experience Sampling, questions are deliberatively phrased in a neutral way so as to avoid pressures on the participants and the time when questions are asked is planned to reduce or avoid memory loss (Hurlburt and Heavey 2001).

The control variables we were able to collect were limited to age, gender, weight, sportiness, personality profiles, weekend vs holidays, and GPS coordinates (latitude, longitude and altitude). In the future, we plan to include also context-based variables such as: with whom individuals are spending their time, whether they are working or are retired, sick or healthy, as well as weather conditions which helps control for the correlation between lighting and reported mood. In this initial experiment, we could not classify the locations to see whether specific coordinates corresponded to specific places for each person (home, workplace, university, or library). This could be a proposal for future research, to test the influence/control of specific locations on Happiness and Activation. It would also be interesting to include traits such as innovativeness which has been found to be predictive of individual intention to continue to adopt smartwatch technology (Hong, Lin et al. 2017).

Unfortunately, Pebble was bought in November 2016 by Fitbit, which discontinued the production of their smartwatch, while maintaining the Cloudpebble software development platform. We bought a supply of smartwatches for our experiments, but for future work we might have to port our platform to other types of smartwatches.

Another limitation is represented by the small sample (17 participants) that was non-randomly selected. Therefore, we suggest replicating our experiment on larger randomly selected samples, including people of different ages and coming from more homogeneous groups, controlling for other variables such as job, history of mental illness and marital status.

The application of our method could be useful in the future for an automated more accurate measurement of changes in mood states. We aim at combining our approach with the analysis of the social networks in which individuals are involved and investigate which relationships lead to improved well-being. Our method has also some marketing application, for example to assess customers' reaction to ads and the effect of store layout.

Body sensors have the potential to propel novel applications that go beyond healthcare research, offering real-time sensing and processing that promises to improve and expand human life. While there are already devices available that can assess changes in health conditions in real time through sensors (Hanson, Powell et al. 2009; Patel, Park et al. 2012; Yang 2006), our model offers the opportunity to add the assessment of mood changes and provide insights to the wearer or the care giver about needed corrective actions. New projects aim to design and implement personal data capturing systems to support social and health services (Haring, Banzer et al. 2015). Recently, Mainetti and colleagues (2016) proposed a personal data capturing system to help elderly people deal with mild cognitive impairments and frailty. In their system, data is collected by means of inertial sensors embedded into wearable devices or smartphones, while environment parameters are gathered through wireless sensor networks or sensors included in portable devices.

While in this paper we measured happiness through parameters related to the body state (e.g. acceleration, heartrate), in the future we plan to correlate Happiness and Activation to relational indicators (e.g. where you have been, and with whom you have been). This is supported by research by Aaker, Rudd et al. (2011) who found that the two main happiness principles were to spend time with the right people, and to spend time on the right activities as demonstrated by the Harvard Grant Study (Vaillant 2012). Our effort is aligned with positive psychology literature focused on affect not only as an intra-psychical trait, but looking also at its social component. Sharing emotions and affect could influence the development of group interactions (Barsade and Gibson 2012). As demonstrated by Christakis and Fowler (2013) in their study on friendship, family, spousal, neighbor, and coworker relationships, people who are surrounded by several happy people - who are also central in their social networks - are more likely to be happy in the future. Their study is particularly important as it shows how happiness spreads across a diverse array of social ties. Building on these studies, we are currently developing an extension of the Happimeter application to implement a comparison of individual sensor and mood data to that of their peers. In prior studies, users have valued such mirroring feedback, and it increased their motivation to continue participating (Gloor, Oster et al. 2010).

Aristotle said "happiness is a state of activity". As we found in this project, happy people are more active. Making activity and happiness more obvious to the users of our system might make them more active, and thus increase their happiness.

## 9. Acknowledgment

The authors are grateful to the referees for their constructive input. The authors would like to thank Cihan Öcal, Jan Marc Misselich, Ines Rieger, Anne Schleicher, Xia Yu, Liqi, and Gavin who, as participants in a collaborative innovation networks seminar at University of Cologne, University of Bamberg, and Jilin University, Changchun, were part of the team building the basic smartwatch software infrastructure. We would also like to thank the team at Swissnex Boston, Dan Aufseesser, Francesco Bortoluzzi, Jonas Brunschwig, Felix Moesner, Sophie Sithamma, Anita Suter, Cécile Vulliemin, Gary Weckx for being early testers of our system. This project has been supported by a series of grants by Philips Lighting (now Signify).